# Green-aware Mobile Edge Computing for IoT: Challenges, Solutions and Future Directions

**Minxian Xu, Chengxi Gao, Shashikant Ilager, Huaming Wu, Chengzhong Xu, Rajkumar Buyya**

**Abstract** The development of Internet of Things (IoT) technology enables the rapid growth of connected smart devices and mobile applications. However, due to the constrained resources and limited battery capacity, there are bottlenecks when utilizing the smart devices. Mobile edge computing (MEC) offers an attractive paradigm to handle this challenge. In this work, we concentrate on the MEC application for IoT and deal with the energy saving objective via offloading workloads between cloud and edge. In this regard, we firstly identify the energy-related challenges in MEC. Then we present a green-aware framework for MEC to address the energy-related challenges, and provide a generic model formulation for the green MEC. We also discuss some state-of-the-art workloads offloading approaches to achieve green IoT and compare them in comprehensive perspectives. Finally, some future research directions related to energy efficiency in MEC are given.

**Keywords** Mobile Edge Computing, Smart Devices, Energy Efficiency, Low Latency, Workloads Offloading

## 1. Introduction

The concept of the IoT has evolved remarkably based on the evolution of wireless communication and mobile technologies [1]. IoT has been regarded as a global network consisting of connected smart devices, which contributes to the arising and evolving of various novel mobile applications. Furthermore, with fast development, IoT has the potential to promote many more possible applications and scenarios, such as smart cities, smart home, smart health-care, smart agriculture, and so on. However, since IoT devices have the inherent features, including constrained power capacity, low computation capacity, and storage, provisioning limited resources for a great amount of computation-intensive applications on devices is a significant challenge [2] [3] [4] [5].

To handle the fast increase of mobile applications and ensure the performance of applications on IoT devices, application tasks are offloaded to cloud that gathers adequate computation resources from remote servers [6]. This motivates the paradigm named mobile cloud computing (MCC) [7]. In MCC, mobile devices can utilize the computing and storage resources from remote clouds, which can be accessed via a core network. The MCC paradigm can extend battery life, enhance mobile devices' capacity to handle complex tasks, provide larger storage space. However, communication cost and service delay are two significant issues that can undermine the user experience, due to increased load on the core network.

To address the above limitation of MCC, Mobile Edge Computing (MEC) paradigm was proposed that enables efficient execution of applications requiring low latency with constrained energy [8]. MEC is a type of computing paradigm that enables capabilities of cloud computing to be extended at the edge of network [9]. However, ensuring low latency as required is still quite challenging, especially when "Internet of Everything" has evolved as a reality based on the recent IoT technologies, and amidst IoT devices are more diverse in their capabilities and requirements. Moreover, energy efficiency has become an extremely important factor in designing MEC solutions as IoT devices have limited energy and battery life. Therefore, without proper coordination among the resource constrained smart IoT devices and offloading necessary tasks to MEC may lead to higher energy costs and latency.

Another way to relieve the energy constraint of MEC enabled IoT system is by utilizing green energy (e.g. solar, wind, etc.) [10]. Using green energy as the energy sources rather than coal-based brown energy alone can reduce the carbon emission efficiently. Besides, the outdoor IoT devices powered by green energy can also extend their battery life. Enabling the edge servers and IoT devices to be supported by green energy reduces the dependency on coal-based energy sources. There have been some proposed green-aware approaches for MEC enabled IoTs [11] [12]. To some extent, the availability of green energy is intermittent and unpredictable, therefore, it is required to design a hybrid power supply of both green energy and brown energy to fully assure the stability and availability of services.

## *1.1 MEC Characteristics*

To support the sustainable development of IoT technology, MEC has been applied to many IoT scenarios. The MEC provides the following useful characteristics:

**Proximity:** Unlike remote clouds, in MEC, edge servers are deployed at the network edge close to the IoT devices. It can be used to process the key data generated by the IoT devices with shorter processing time as edge servers are generally more powerful than IoT devices.

**Low Delay:** Offloading data from IoT devices to edge servers can achieve low delay (e.g. data transmission time, task processing time), improve user experience, and reduce potential core-network congestion. It is also possible to support real-time applications for time-critical emergency IoT applications.

**High Bandwidth:** The communications between IoT devices and MEC servers can fully utilize the available bandwidth and gain high transmission rate, which can improve the system performance of MEC-enabled IoT.

**Location and Mobility Awareness:** With real-time location data received from IoT devices, the application can estimate the status of the whole system. In addition, in case of mobile devices that move dynamically, tasks can be offloaded to a set of proximal MEC servers. There is a requirement for continuous task offloading service that is seamlessly integrated with platforms.

**Flexible Deployment:** MEC is able to host critical missions with IoT applications. These applications can be deployed by the network managers or third-party developers rather than only from cloud service providers.

**Heterogeneous Resource Collaboration:** To handle the large number of computing workloads, the services require to utilize resources both from cloud computing and edge computing together. Furthermore, coordinating heterogeneous resources is also required to meet the different requirements of various applications.

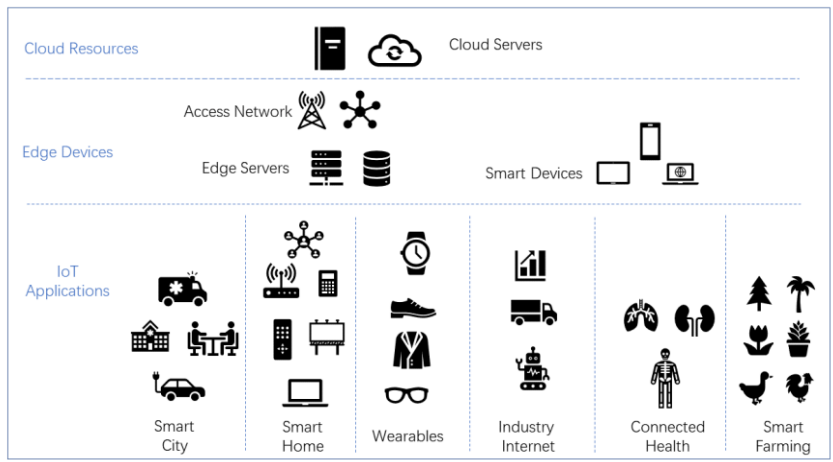

**Fig. 1. MEC for Typical IoT Applications**

Figure 1 shows the MEC enabled IoT scenario. Based on the above features, MEC has been applied to many areas and consists of different types of IoT applications, e.g. transportation, smart grid, agriculture, and healthcare. These IoT applications along with their IoT devices can be placed at the edge of the network. To be more specific, the edge devices can be deployed at the network gateway, base stations, or local area network, which can connect the IoT devices via 5G or WiFi. As for the cloud resources they work as a central manager and monitor the status of edge devices. They can also act as remote repositories to store the data and perform off-line batch processing tasks.

## *1.2 Need for Sustainable IoT Application Management in MEC*

According to the above discussions, an efficient offloading policy is quite important to support the effectiveness of MEC for IoT, as it can adequately allocate resources in an energy-efficient manner while satisfying the latency requirements. To develop efficient offloading policies of MEC enabled IoT, the following challenges should be addressed:

**Heterogeneity of Edge Servers and IoT Devices:** Both the mobile devices and edge servers have heterogeneous network, computing and storage resources, which makes the selection of offloading devices a challenge. For instance, some of the edge servers are suitable for processing compute-intensive tasks while some others are with adequate storage resources. When designing offloading policies, the heterogeneity of resources should be considered to take full advantage of resources.

**Offloading Trigger:** When to trigger the offloading process should be carefully investigated. Always offloading tasks without context-awareness to edge servers can lead to higher delay, if the communication cost is much higher than the processing cost on edge servers in some particular situations, e.g. network congestion, can lead to degraded performance. Therefore, an efficient trigger mechanism is essential.

**Coordination Costs:** The communication cost exists among devices to coordinate tasks, e.g. mobile games. Thus, coordination among devices may consume extra energy and incurs additional latency because of communication overhead. Furthermore, the costs grow exponentially with the increased amount of devices, thus the bottleneck exists when scaling the number of IoT devices to a large scale.

**Partial Task Offloading:** Apart from full task offloading, tasks can be partitioned into different parts. Thus, how to select the appropriate parts to offload to ensure the latency and energy requirement is another challenge, especially for the cases when there is a data dependency between different parts.

**Security Guarantee:** In the IoT network, MEC servers can encounter security attacks like the masquerade attacks. Privacy information can also be revealed in the offloading process. Protecting privacy information while maintaining operational efficiency is a critical challenge.

In summary, when designing efficient workloads offloading policy in MEC enabled IoT, and to address the above challenges, some key research questions should be considered, including:

- When to offload the task to edge servers?
- Partial offloading or full offloading of application tasks?
- Which edge server should be selected to process the offloaded tasks?

In this work, to address the aforementioned challenges, we present a green-aware framework for MEC. We focus on the problem modeling for the workloads offloading in MEC for IoT. In addition, we review some state-of-the-art green-aware offloading approaches.

The main **contributions** of this work are as follows:

- We propose a green-aware framework to support the MEC enabled IoT by taking advantage of green energy to reduce the power consumption and service latency.
- We model the task offloading approach in a general way by considering the local processing model and edge processing model to achieve an energy-efficient and QoS-aware objective.
- We review state-of-the-art green-aware workloads offloading approaches of MEC-enabled IoT to identify the advantages and limitations of current solutions.
- We outline the future research directions in the related area to help the researchers to investigate the future possible trends.

To help the readers to follow the contents easily, the abbreviation notations used in this work are summarized in Table 1.

**Table 1. Summary of abbreviation notations used in this work**

| Abbreviations | Meaning |
|---|---|
| MEC | Mobile Edge Computing |
| IoT | Internet of Things |
| MCC | Mobile Cloud Computing |
| GS-MEC | Green and Sustainable Mobile Edge Computing approach |
| LSTM | Long Short-Term Memory |
| LSDQN | Long Short-Term Memory enhanced Deep Q-Network approach |
| DQN | Deep Q-Network |
| LETOC | Lyapunov-based algorithm for online optimization |
| GOLL | Green Offloading with Low Latency |
| SOMEC | A Selective Offloading in Mobile Edge Computing approach |
| GreenEdge | Approach leveraging device-to-device communication and energy harvesting |

The rest of this paper is organized as follows: we start by presenting the general green-aware framework for MEC enabled IoT in Section 2, where we highlight the latest advances and trends in green-aware MEC. In Section 3, we formulate the general problem modeling and the offloading approaches in green-aware MEC enabled IoT. Then we discuss state-of-the-art green-aware approaches for MEC enabled IoT by identifying their merits and limitations in Section 4. Afterward, we present a number of future research directions in Section 5. And final, we conclude the work in Section 6.

## 2. Green-aware Framework for MEC

Green-aware MEC for IoT aims at reducing energy consumption and communication delay, which plays a crucial role in the IoT paradigm by taking advantage of green energy. In the IoT scenario, it is more important to consider the limited energy capacity of IoT devices. Extending the active time of IoT devices can enhance the lifetime, which makes the task offloading necessary to save the power consumption of IoT devices. The battery status of IoT devices can be obtained in a real-time manner and then reserved in an energy buffer when the IoT devices interact with surroundings. However, it is challenging to achieve the energy efficiency goal via task offloading. In this regard, we propose a framework of green-aware MEC enabled IoT. Figure 2 shows our proposed framework and detailed components are introduced as follows:

The major components in the framework can be divided into two parts including IoT devices and edge servers. The main components in the IoT devices part are: Energy Manager, QoS Manager, Offloading Scheduler, and Synchronizer.

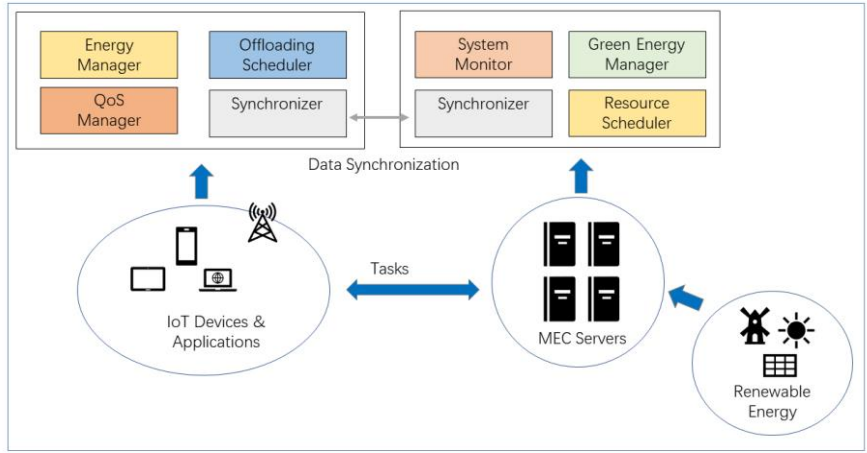

**Fig. 2. Framework of Green-aware MEC for IoT**

**Energy Manager:** it is responsible for managing the energy usage of IoT devices. Based on energy usage, it can also trigger the offloading operations.

**QoS Manager:** it monitors the QoS information of IoT devices and applications, such as communication latency. It also determines the service requirements and anticipated latency for executing the tasks.

**Offloading Scheduler:** it decides whether to offload the tasks or not as well as which part of tasks should be offloaded. It also partitions the tasks and selects the target edge server to send the offloaded task.

**Synchronizer:** it is responsible for handling the communications and synchronizing the data when offloading the tasks, e.g. full or partial offloading. For instance, some data should be processed locally and the processed data in edge servers should be synchronized when data are sent back. It interacts with the corresponding Synchronizer component in the edge server part, and it also ensures data integrity.

In our proposed framework, the edge servers can be powered by both brown energy and green energy. The green energy is produced via renewable sources, like solar and wind. The major components in edge servers part include System Monitor, Green Energy Manager, Resource Scheduler, and Synchronizer.

**System Monitor:** it monitors the status of edge servers, including CPU usage and storage usage. It can also alert the anomaly of the system.

**Green Energy Manager:** it controls the green energy usage for edge servers in terms of the availability of green energy to maximize the utilization of green energy. It can also include the green energy prediction module.

**Resource Scheduler:** it manages the resources in MEC devices to support the process of offloaded tasks to reduce the processing time by allocating resources to corresponding tasks.

**Synchronizer:** it is responsible for synchronizing the data with IoT devices. It receives the tasks and sends the data back to IoT devices and ensures them to be consistent.

## 3 Problem Modelling: Green-Aware Offloading

Based on our proposed framework, our target problem of MEC for IoT can be modeled in the following way, which contains the task model, green energy model, local processing model, and edge processing model.

### 3.1 Task Model

In the whole system, we assume that there are $M$ edge severs, the processing capacity of the servers can be represented as $f_k^{max}$, where $k = 1,2,...,M$. These edge servers are deployed independently in $M$ base stations located in different areas. Considering there are $N$ IoT devices and $N(t)$ active IoT devices in the system at time interval $t$ among the whole scheduling period $T$. Each device has a compute-intensive task, while can be denoted as $T_i = \{s_i, c_i, d_i\}$, where $i = 1,2,...,N$, $s_i$ is denoted as the data amount of the input task, $c_i$ is the required computation resources, e.g. millions of instructions. $d_i$ is the deadline constraint of this task. For $T_i$, it can be consist of three parts: $x_i^l(t)$, $x_i^e(t)$, and $x_i^d(t)$. The $x_i^l(t)$, $x_i^e(t)$ denote the percentage (should be a real value between 0 and 1) of tasks processed on IoT devices locally or tasks executed on edge servers respectively. The $x_i^d(t)$ denotes the admission control by dropping tasks, which is either 0 or 1, representing whether the task is dropped or not. These three parameters should conform to the following constraint:

$$x_i^l(t) + x_i^e(t) + x_i^d(t) = 1, \forall t \in T, \forall i \in N \tag{1}$$

$$x_i^l(t), x_i^e(t) \in [0,1], x_i^d(t) \in \{0,1\}, \forall t \in T, \forall i \in N \tag{2}$$

## *3.2 Green Energy Provisioning Model*

The availability of green energy can vary significantly in different locations with varied weather conditions [13]. For example, in some locations with the summer time, the solar power is adequate, while in some other places with the winter time, the wind can be the main green energy sources. In addition, the availability is heavily dependent on the weather conditions, therefore, can vary significantly in different time zones. In our model, the green energy has a higher priority to be used, which means the green energy will be used firstly until it lasts and the coal-based brown energy will be used as the complementary.

We consider that the edge servers are powered by both brown energy and green energy. At time interval $t$, the amount of green energy is $R_k(t)$, which is tightly coupled with the available amount of renewable energy, for instance, the solar power is 0 at night while it can reach the peak at noontime in a sunny day. And $k$ denotes the location that offers green energy. To make the system extensible, we also consider that there is a set of cloud servers behind edge servers as backups.

## *3.3 Local Processing Model*

In the local processing model, the tasks are processed locally, thus $x_i^l(t) = 1$, $x_i^e(t) = 0$ and $x_i^d(t) = 0$. Assuming the local IoT device processing capacity is $f_i(t)$ at time interval $t$, which is constrained by the maximum processing capacity $f_i^{max}$ as $f_i(t) \leq f_i^{max}$. Then the task processing delay $D_i$ for executing $T_i$ is as follows:

$$D_i = \frac{c_i}{f_i(t)}, \quad \forall t \in T, \forall i \in N \tag{3}$$

Let $P_i^l(t)$ represent the energy consumption of processing the tasks locally. Derived from [14], the energy consumed for the task by local processing can be expressed as:

$$P_i^l(t) = k \cdot \big(f_i(t)\big)^2 \cdot D_i, \quad \forall t \in T \tag{4}$$

where $k$ is the energy factor depends on the hardware architecture.

Green-aware MEC for IoT: Challenges, Solutions and Future Directions    9

## 3.4 Edge Processing Model

In contrast to the local processing choice if the task $T_i$ is offloaded to edge server $k$, the total processing time will be constituted with 3 parts, including the running time for device to access edge server $k$, the task communication time and the processing time. In this way, $x_i^l(t) = 0$, $x_i^e(t) = 1$ and $x_i^d(t) = 0$. Therefore, the time delay $D_{i,k}$ of processing the task $T_i$ on the edge server $k$ can be represented as:

$$D_{i,k} = \frac{d_i}{r_k} + \frac{c_j}{f_{i,k}(t)} + C_k \tag{5}$$

where $r_k$ is the device transmission rate to access base station $k$, based on Shannon-Hartley formula, the transmission rate from the IoT device to edge server can be calculated as $r_k = w log_2(1 + \frac{s \cdot p(t)}{\sigma})$, where $w$ and $\sigma$ are the channel bandwidth and noise power, $s$ is the transmission power of IoT device, and $p(t)$ is the channel gain from the device to the server. $f_{i,k}(t)$ is amount of the communication resources allocated by MEC server $k$ at time interval $t$ for offloading $T_i$. $C_k$ is the running time for device to connect the MEC server $k$, which can be a value in a range, e.g. 5 to 50 ms.

Let $P_i^e(t)$ denote the power consumption of executing the tasks on the edge servers. Therefore, the energy used for processing the tasks on the edge servers can be represented as:

$$P_i^e(t) = p_i \cdot D_{i,k}, \qquad \forall t \in T \tag{6}$$

where $p_i$ is the transmission power scheduled for IoT device $i$.

## 3.5 Optimal Green-aware Offloading

Let us denote $R_i$ as the reward of device $i$ gains by offloading the task $T_i$ to the MEC server $k$. The $R_i$ is correlated with the delay reduction and energy consumption improvement. We assume that the scheduler in IoT devices can make efficient decisions, thus the system can maximize their reward by selecting the offloading choices.

$$\max_{T_{i,k}, f_{i,k}} R_i = \sum_{k=1}^{M} T_{i,k} (\lambda(D_{i,0} - D_{i,k}) + \epsilon(P_{i,0} - P_{i,k})) \tag{7}$$

$$s.t. \sum_{k=1}^{M} T_{i,k} \cdot D_{i,k} \leq d_i \tag{8}$$

$$\sum_{k=1}^{M} T_{i,k} f_{i,k} \leq f_k + f_k^b \tag{9}$$

$$constraints\ (1), (2) \tag{10}$$

where $D_{i,0}$ and $P_{i,0}$ are the time and energy consumption to locally process the task on IoT devices. $P_{i,k}$ is the energy cost for offloading task $T_i$ to MEC server $k$.

The $\lambda$ and $\epsilon$ are coefficients. The $f_k^b$ represents the computing resource used from the backup server MEC server $k$. Some system constraints should also comply like the maximum green energy usage should be more than the capacity of green equipment. The offloaded tasks should not surpass the maximum allowed tasks. The choices are two-folded: offloaded or not. The CPU utilization of offloaded tasks should not be more than the maximum resource utilization.

Once the tasks are offloaded, MEC servers optimize their resource usage by taking advantage of green energy. Since renewable energy availability heavily dependent on the many factors related to weather, using green energy is one of the motivated policies of the MEC servers. Therefore, for the offloading process, in the non-cooperative game between the MEC servers, the possible decisions of server $k$ can be denoted as $(x_k, f_k^b)$. For edge server $k$, then the optimization problem is formulated as:

$$\max_{x_i, f_k^b} R_k^s = x_k \sum_{i=1}^{N} f_{i,k} - \beta_k \cdot \min\left(\sum_{i=1}^{N} f_{i,k}, f_k^{max}\right) - y \cdot f_k^b \qquad (11)$$

$$s.t.\ x_k, f_k^b \geq 0 \qquad (12)$$

where $\beta_k$ is the changing rate/transformation rate of green energy at location $k$ that can be computed based on the green energy model in Section 3.2, and $y$ is the ratio of resources bought from backup servers.

Since the computation capacity of each MEC server is limited, there is competition among offloaded tasks to utilize the resources, which makes the decisions to be a non-cooperative game. The task offloading process among devices is a concave multiple-player game, thus a Nash equilibrium can exist [15]. As the reward function $R_k^s$ continues in terms of $x_k$ and $f_k^b$, it can be easily solved by integer programming, Lyapunov optimization, game theory, or other approaches.

**Scalability discussion:** the scalability of the proposed work depends on the methods that solve the above problem. For example, Lyapunov optimization based algorithms can be more time-consuming than heuristic algorithms when the number of devices in the system increases largely, while Lyapunov optimization based algorithms can achieve better optimized results. Therefore, there are trade-offs between the performance and scalability that can be determined by the service providers according to their focus.

## 4 State-of-the-art Offloading Approaches

In the following, we provide an overview of existing approaches to green-aware MEC enabled IoT to identify their advantages and limitations. We also present a comparison among the investigated approaches from multiple perspectives. Although there are numerous research works addressing offloading in MEC, we focus on essential works that directly address the green-aware offloading problem.

## 4.1 GS-MEC

Green and Sustainable Mobile Edge Computing (GS-MEC) [11] is a framework to support IoT devices to be self-powered by taking advantage of green energy for the IoT scenario. Its optimization objective is to improve energy efficiency and system sustainability. Compare with the traditional communication framework for the Cloud enabled IoT environment, GS-MEC adopts a parallel offloading strategy in MEC. GS-MEC considers packet losses to ensure the reliability of the framework. Energy Harvesting Technologies are applied to take full usage of green energy supporting the IoT system in a smart home, which is equipped with a set of IoT devices. To power the devices with green energy, the IoT devices are consist of energy-harvesting components that can convert green energy into electrical energy. Based on the models including computing model, energy model, task cost, the optimization problem is formulated as a minimization problem of latency and maximization task admission rate under the constraint of energy. An offloading algorithm based on Lyapunov optimization [16] is also proposed in GS-MEC to solve the optimization problem. Lyapunov optimization is used to decompose the formulated problem into subproblems that can be solved easier and apply the variable substitution optimization technique to decompose variables. In the proposed algorithm, the energy consumption of IoT devices, transmission power, CPU frequency of IoT devices, and the offloading decision would be obtained in each time interval. And it can offer sufficient power for IoT devices by controlling both the CPU frequencies and transmission power by managing energy.

Both theoretic analysis and simulation results have demonstrated the proposed approach can reduce latency efficiently. The advantage of the framework is that it considers the parallel offloading. However, the proposed algorithm is only compared with some simple baselines.

## 4.2 LSDQN

LSDQN [12] is an approach based on Long Short-Term Memory (LSTM) enhanced Deep Q-Network for dynamic task management problem in MEC. Its objective is to improve the average uplink transmission rate while reducing the power consumption of the IoT network. LSDQN considers that IoT devices can be supported by the power from a rechargeable battery via harvesting energy from the nearby environment. In the proposed approach, LSTM is applied to estimate the battery status to provide information for IoT devices to make access control based on Deep Q-Network (DQN). LSTM is a widely used network structure of recurrent neural networks to solve the gradient disappearance problem via storing historical information in memory, and DQN is a machine learning method that can learn the optimal policy as per optimization function. To achieve the objective, the uplink rate and energy optimization problem is modelled as a Markov decision process without the knowledge of system dynamics. A MEC access control management policy is also proposed. In the policy, the dynamics of energy and network status are considered to support the decision for the tasks about which MEC server to offload the tasks to. The proposed approach can be applied to the scenario with limited information about future energy supply by taking advantage of LSTM to reduce the prediction loss. In each time interval, the IoT device chooses the most suitable MEC server to transfer data based on the device state and battery status. A software-defined network manager is also applied to capture all the system status. Experiments based on simulations have demonstrated the effectiveness of the proposed approach.

### *4.3 LETOC*

LETOC [17] is a Lyapunov-based algorithm for online optimization on energy cost and time to address the energy-aware computation offloading in MEC for IoT. The objective of LETOC is to minimize the long-term cost of system while ensuring the user experience in terms of quality of service. LETOC is a near-optimal policy for deciding control actions on application offloading by balancing the trade-offs between cost and response time. It also considers taking advantage of green energy, thus reducing the usage of brown energy. LETOC incorporates green energy sources to ensure the IoT devices are running well and manages energy-efficient data offloading. The optimization problem is converted into a constrained stochastic optimization problem and then solved based on Lyapunov optimization. LETOC can distribute incoming tasks to the corresponding servers without prior knowledge of user and system status. LETOC also takes into account the cost related to fees paid for grid power. A discrete time-model is considered in the scheduling process. Optimality analysis is also provided for LETOC, which proves the proposed approach can achieve the average response to be close to the theoretical optimum. Simulation evaluations have validated the performance of the proposed approach, which achieves better performance than the baselines.

### *4.4 GreenEdge*

GreenEdge [18] is an approach leveraging device-to-device communication and energy harvesting techniques to support task execution in a sustainable and collaborative manner. Device to device communication is defined as the direct communication between two wireless devices in proximity by passing information through the base station. GreenEdge aims to reduce the power demand of IoT devices via offloading more workloads to devices that support energy harvesting, especially for the situation when IoT devices have insufficient energy supply. Tasks in GreenEdge can be executed in three ways: local execution, device to device offloaded execution, and edge offloaded execution. The discrete time-slotted model is adopted to capture the system dynamics like in Section 4.3. According to current task characteristics and renewable energy availability, in each time interval, the resource scheduler in GreenEdge dynamically optimizes the execution mode for each task to optimize the energy efficiency of edge devices. The optimization problem is formulated as a minimizing problem of the long-term brown energy usage by managing the rechargeable battery usage at each time interval. Then the optimization problem is solved by the Lyapunov-based online optimization framework. The optimization can also make trade-offs between brown energy and battery energy usage. The proposed approach has shown the possibility to be applied to some IoT applications, e.g. smart street lighting and smart bike-sharing.

## 4.5 GOLL

Green Offloading with Low Latency (GOLL) [19] is a mobility-aware and layered MEC framework to support low-latency and green IoT. It aims to exploit MEC for mobile devices to support multiple layer computing resources to be shared among edge servers. During the task offloading process, the task offloading policy selected by the smart devices is determined based on the computing price of edge servers. GOLL considers that edge servers can be indirectly connected according to their offered resource prices during the service bidding process, in which the edge server with the lowest resource price can be offloaded with tasks. Therefore, the price is considered as the link to two non-cooperative games among the smart devices and MEC servers. In this work, a Stackelberg game [20] is also applied to handle the offloading problem in the proposed framework, which is a solution for the optimal offloading problem. In the proposed framework, the utilization of MEC servers can be improved while the energy consumption and service latency can be reduced. An iterative-based heuristic algorithm is utilized to achieve the corresponding results. In each iteration, the smart device can respond as per the prices announced by edge servers, and the edge servers can make the optimal decision according to the obtained response of the devices. When the strategy is not updating anymore, the iteration process stops. The efficiency of the proposed schemes is validated through numerical results, which shows better performance than baselines.

## 4.6 SOMEC

A selective offloading in mobile edge computing (SOMEC) is proposed in [21] for green Internet of Things, which is included in a lightweight framework to deal with the scalability problem. The approach does not need the coordination among devices and can operate at the IoT device and edge servers separately by integrating latency constraints in requests. The objective of the SOMEC is minimizing the power consumption of devices. The communication overheads can also be reduced by device self-nomination for tasks processing or self-denial for tasks. The proposed framework is lightweight concerning communication overheads. The devices can independently send offloading requests and the servers can make decisions on whether to admit requests or not. The working process of the proposed framework mainly contains three steps. Firstly, each mobile device will dispatch an offloading request to the selected edge server including the resource and QoS requirements. Thereafter, servers receive the offloading requests, and they would only admit some of the selected users or workloads for offloading. In this step, the corresponding resources should be allocated to meet the requirements. Finally, the mobile devices can offload the tasks based on the admission results.

## *4.7 Discussions of the Investigated Work*

To compare the differences among the investigated papers, we compare these works from multiple perspectives as shown in Table 2. The different perspectives include:

**Environment:** It represents the environment that the proposed approach can be applied, including a heterogeneous or homogeneous environment, which manages heterogeneous or homogeneous resources and devices. In the investigated approaches, most of them are targeting for the heterogeneous environment except for LETOC, which is applied for the homogeneous environment.

**Optimization Objective:** It is the primary objective that the investigated paper aims to achieve. As green-aware scheduling is one of the focus of these works, energy-related optimization is the objective shared by all investigated papers. However, these papers have some differences in other optimization objectives. For instance, GS-MEC focuses on minimizing response time and packet loss, while LSDQN pays more attention to transmission rate optimization. LETOC aims to balance the response time and energy cost, and GreenEdge targets to minimize fossil energy usage while ensuring task performance. GOLL focuses on maximizing service provider utilities, while SOMEC spends more effort on reducing communication overheads.

**Energy-Saving Component:** It represents the component that will optimize energy by the investigated approaches. The energy consumption optimization of IoT devices is managed by GS-MEC, LETOC, GOLL, and SOMEC. As for GreenEdge, it provides a holistic energy optimization for edge data centers. LSDQN focuses on optimizing the energy of the IoT network.

**Green Energy Sources:** It is the energy sources that provide green energy to support the green-aware scheduling of the system. GS-MEC and GreenEdge utilize energy harvester, LSDQN uses a rechargeable battery, LETOC considers solar energy, GOLL and SOMEC consider green energy from both solar and wind.

**Workloads:** It compares the workloads used in different approaches for performance evaluations. Synthetic workloads are used in LSDQN, LETOC, and GOLL. GS-MEC evaluates performance with image compression application, and SOMEC analyzes with face recognition applications. As for GreenEdge, it applies commercial IoT application workloads.

Table 2. Comparison of state-of-the-art approaches

| Approach | Environment | Optimization Objective | Energy-saving Component | Green Energy Sources | Workloads | Experiments Platform | Merits | Demerits |
|---|---|---|---|---|---|---|---|---|
| GS-MEC [11] | Heterogeneous | To minimize response time and packet loss under energy limitation | IoT devices | Energy harvester | Images compression application | Simulation | Reduced completion time, task cost, and ratio of dropped task | Performance under heterogeneous environment can be further investigated |
| LSDQN [12] | Heterogeneous | To improve uplink transmission rate while minimizing transmission energy | IoT network | Rechargeable battery | Synthetic | Simulation | Improved uplink rate and reduced energy consumption | Weather forecasting approach can be improved |
| LETOC [17] | Homogeneous | To balance response time and energy cost | IoT devices | Solar energy | Poisson distribution service rate | Simulation | Near optimal solution for balancing response time and energy cost | Weather forecasting approach can be improved |
| GreenEdge [18] | Heterogeneous | To minimize fossil energy consumption while ensuring task performance | Edge data centers | Energy harvester | Commercial IoT applications | Theoretical analysis | Device to device communication is considered | Performance evaluations should be conducted |
| GOLL [19] | Heterogeneous | To maximize service provider utilities while reducing device energy consumption | IoT devices | Solar, wind | Synthetic | Numerical analysis | Reduced latency and energy | More parameters can be evaluated in experiments, e.g. varied vehicle speed |
| SOMEC [21] | Heterogeneous | To minimize energy consumption and reduce communication overheads | IoT devices | Solar, wind | Face recognition application | Numerical analysis | Improved system scalability | Green usage can be added into the decision engine component |

**Experiments Platform:** It represents the experimental approach for evaluating the performance as well as the platform for conducting experiments. GS-MEC, LSDQN, and LETOC are using simulations, GreenEdge applies theoretic analysis, GOLL and SOMEC conduct numerical analysis.

**Merits and Demerits:** It summarizes the advantages and disadvantages of investigated papers. For example, MS-MEC can reduce completion time, task cost, and the ratio of the dropped task, however, the performance under a heterogeneous environment should be further investigated.

## 5 Future Research Directions

As discussed in the previous sections, the green-aware offloading for MEC enabled IoT has attracted attention and achieved significant progress in recent years benefiting from its ability to reduce energy while ensuring system performance. However, there are some research challenges that should be further explored to make MEC more efficient and reliable. This section discusses several future research directions outlining the different avenues.

**Energy Consumption Optimization:** Current works mostly focus on the management of the energy consumption from edge servers. More energy consuming parts, e.g. network energy consumption can be further explored.

**Evaluations with Real Testbed:** The experiments of current research are mostly based on simulations or numerical analysis. There is a lack of a real testbed or prototype system that can evaluate the performance of proposed approaches.

**Benchmarks:** Presently, no standard benchmarks for performance evaluation of green-aware offloading approaches are provided. A benchmark is demanded to evaluate the energy efficiency performance of novel algorithms and compare them with other approaches aiming for similar objectives.

**Collaboration with MCC:** MEC has been validated as an effective way to reduce the latency of services and improve user experiences. However, it is still promising to take the heterogeneity of both mobile cloud computing and mobile edge computing together to build a hybrid environment for selecting task offloading destinations. Furthermore, considering distributed cloud data centers can also improve the usage of green energy.

**Green Energy Usage Maximization:** The availability of green energy keeps changing along with time. How to take advantage of green energy to support IoT devices can be further investigated, e.g. machine learning or deep learning approaches to predict the availability of green energy and resource usage.

**Varied QoS Satisfactory:** Some of the current research has considered the trade-offs between energy consumption and quality of service. This can be further investigated by considering the optimal way to physically allocate the resources according to expected users with varied QoS requirements rather than a single QoS requirement.

**Managing the Mobility of IoT Devices:** Most of the research propose the offloading decision that assumes strictly static scenarios rather than dynamic scenarios, i.e. the IoT devices do not move during the offloading time. However, the transmission rate can be significantly influenced if channel quality drops during the movement. This can lead to more energy consumption or higher delay. Therefore, a more advanced model considering the movement of IoT devices should be proposed. Predictions techniques for movement can be explored.

**Joint Data Management:** Current research focuses on offloaded data while neglecting the conventional data that is not offloaded to the MEC, e.g. HTTP, FTP that has to be transmitted over backhaul links and radio in parallel to the offloaded data. Therefore, it is required to schedule communication resources for the management of conventional data (e.g. the data not exploiting MEC). Therefore, the joint data management approach for both offloading and conventional data is required.

**Multi-tenancy Management:** The MEC computing infrastructure is shared environment and different user's applications will be hosted on MEC servers. Hence, solving the inference issues and providing the performance isolation for applications to guarantee the required SLAs is crucial.

**Security Management:** The security is an important aspect requiring solutions across the computing and network stack. In specific to MEC, it is more challenging to provide the privacy for user data due to shared resources and continuous streaming data that flows between IoTs and MEC servers [22]. Considering the resource capabilities in IoTs and MECs, light weight security solutions need to be incorporated to manage the privacy and confidentiality of the user's or application's data.

## 6 Summary and Conclusions

In this work, we present a discussion on green-aware mobile edge computing for IoT. Specially, we discuss the related challenges about how to apply MEC for IoT to achieve the energy efficiency objective. Moreover, we propose a general framework including the necessary entities to support the green-aware resource scheduling in the MEC scenario. Thereafter, we present a green-aware model for offloading tasks from IoT devices to edge servers to achieve the efficient management of energy and latency. Then we investigate several state-of-the-art approaches in the related area and compare them from comprehensive perspectives. Finally, we provide a set of future research directions, where we hope to attract researchers' attention to establish more validated research in the green-aware MEC enabled IoT area, e.g. collaborating the MEC with MCC together to take advantage of the heterogeneity of them for task offloading.

**Acknowledgements**: This work is supported by Key-Area Research and Development Program of Guangdong Province (NO. 2020B010164003), and SIAT Innovation Program for Excellent Young Researchers.

# References

1. Redowan Mahmud, Ramamohanarao Kotagiri, and Rajkumar Buyya. Fog Computing: A Taxonomy, Survey and Future Directions, pages 103-130. Springer Singapore, Singapore, 2018.
2. Nirwan Ansari and Xiang Sun. Mobile edge computing empowers internet of things. IEICE Transactions on Communications, 101(3):604-619, 2018.
3. Yi Liu, Chao Yang, Li Jiang, Shengli Xie, and Yan Zhang. Intelligent edge computing for iot-based energy management in smart cities. IEEE Network, 33(2):111-117, 2019.
4. Pavel Mach and Zdenek Becvar. Mobile edge computing: A survey on architecture and computation offloading. IEEE Communications Surveys & Tutorials, 19(3):1628-1656, 2017.
5. Luigi Atzori, Antonio Iera, and Giacomo Morabito. The internet of things: A survey. Computer Networks, 54(15):2787-2805, 2010.
6. Minxian Xu, Rajkumar Buyya. BrownoutCon: A software system based on brownout and containers for energy-efficient cloud computing. Journal of Systems and Software, 155:91-103, 2019.
7. Niroshinie Fernando, Seng W Loke, and Wenny Rahayu. Mobile cloud computing: A survey. Future generation computer systems, 29(1):84-106, 2013.
8. H. Wu, W. J. Knottenbelt, and K. Wolter. An efficient application partitioning algorithm in mobile environments. IEEE Transactions on Parallel and Distributed Systems, 30(7):1464-1480, July 2019.
9. Shinan Song Zhanyang Zhang Chengxi Gao Shuhui Chu, Zhiyi Fang and Chengzhong Xu. Efficient Multi-Channel Computation Offloading for Mobile Edge Computing: A Game-Theoretic Approach. IEEE Transactions on Cloud Computing, pages 1-12, 2020.
10. Minxian Xu, Adel N. Toosi, Behrooz Bahrani, Reza Razzaghi, and Martin Singh. Optimized renewable energy use in green cloud data centers. In Sami Yangui, Ismael Bouassida Rodriguez, Khalil Drira, and Zahir Tari, editors, Service-Oriented Computing, pages 314-330, Cham, 2019. Springer International Publishing.
11. Yiqin Deng, Zhigang Chen, Xin Yao, Shahzad Hassan, and Ali MA Ibrahim. Parallel offloading in green and sustainable mobile edge computing for delay-constrained iot system. IEEE Transactions on Vehicular Technology, 68(12):12202-12214, 2019.
12. Lijuan Xu, Meng Qin, Qinghai Yang, and KyungSup Kwak. Deep reinforcement learning for dynamic access control with battery prediction for mobile-edge computing in green iot networks. In 2019 11th International Conference on Wireless Communications and Signal Processing (WCSP), pages 1-6. IEEE, 2019.
13. Minxian Xu and Rajkumar Buyya. Managing renewable energy and carbon footprint in multi-cloud computing environments. Journal of Parallel and Distributed Computing, 135:191-202, 2020.
14. Nikzad Babaii Rizvandi, Javid Taheri, and Albert Y. Zomaya. Some observations on optimal frequency selection in dvfs-based energy consumption minimization. Journal of Parallel and Distributed Computing, 71(8):1154-1164, 2011.
15. Robert Aumann and Adam Brandenburger. Epistemic conditions for nash equilibrium. Econometrica, 63(5):1161-1180, 1995.
16. Lei Zheng and Lin Cai. A distributed demand response control strategy using Lyapunov optimization. IEEE Transactions on Smart Grid, 5(4):2075-2083, 2014.
17. Yucen Nan, Wei Li, Wei Bao, Flavia C Delicato, Paulo F Pires, Yong Dou, and Albert Y Zomaya. Adaptive energy-aware computation offloading for cloud of things systems. IEEE Access, 5:23947-23957, 2017.
18. Zhi Zhou. Greenedge: Greening edge datacenters with energy-harvesting iot devices. In 2019 IEEE 27th International Conference on Network Protocols (ICNP), pages 1-6. IEEE, 2019.
19. Ke Zhang, Supeng Leng, Yejun He, Sabita Maharjan, and Yan Zhang. Mobile edge computing and networking for green and low-latency internet of things. IEEE Communications Magazine, 56(5):39-45, 2018.

20. Jin Zhang and Qian Zhang. Stackelberg game for utility-based cooperative cognitive radio networks. In Proceedings of the tenth ACM international symposium on Mobile ad hoc networking and computing, pages 23-32, 2009.
21. Xinchen Lyu, Hui Tian, Li Jiang, Alexey Vinel, Sabita Maharjan, Stein Gjessing, and Yan Zhang. Selective offloading in mobile edge computing for the green internet of things. IEEE Network, 32(1):54-60, 2018.
22. Pardis Emami Naeini, Sruti Bhagavatula, Hana Habib, Martin Degeling, Lujo Bauer, Lorrie Faith Cranor, and Norman Sadeh. Privacy expectations and preferences in an iot world. In Thirteenth Symposium on Usable Privacy and Security (SOUPS 2017), pages 399-412, 2017.